\documentclass[12pt]{iopart}
\usepackage{tensor}
\usepackage{iopams}
\usepackage{float}
\usepackage{tikz}
\usepackage{graphicx}
\usepackage{float} % For [H] positioning
\usepackage{subcaption}
\usepackage[normalem]{ulem}
\usetikzlibrary{backgrounds}
\usepackage{subcaption}
\newcommand{\eqref}[1]{(\ref{#1})}
\newcommand{\dd}{\mathrm{d}}

\newcommand{\pdv}[2]{\frac{\partial #1}{\partial #2}}

\newcommand{\V}{\mathcal{V}}
\newcommand{\T}{\mathcal{T}}
\newcommand{\Tgr}{T_{\textrm{\tiny{gr}}}}

\newcommand{\kc}{\kappa_{\textrm{\tiny{C}}}}
\newcommand{\kr}{\kappa_{\textrm{\tiny{R}}}}

\newcommand{\dddot}[1]{\stackrel{...}{#1}}

\begin{document}
\title[Gravitational entropy in Petrov Type I spacetimes]{Gravitational entropy in Petrov Type I spacetimes}
\author{M Sarma$^1$\footnote[7]{Corresponding author.}, S N\'ajera$^2$, R A Sussman$^3$}

\address{$^1$ Aix Marseille Univ, Universit\'e de Toulon, CNRS, CPT, Marseille, France}
\ead{sarma@cpt.univ-mrs.fr}

\address{$^2$ Instituto de Ciencias F\'isicas, Universidad Nacional Aut\'onoma de M\'exico, 62210, Cuernavaca, Morelos.}
\ead{sebastian.najera@icf.unam.mx}

\address{$^3$ Instituto de Ciencias Nucleares, Universidad Nacional Aut\'onoma de M\'exico
(ICN--UNAM),\\ A. P. 70 –- 543, 04510 M\'exico D. F., M\'exico.}
\ead{sussman@nucleares.unam.mx}

\begin{abstract}
The gravitational entropy proposal of Clifton, Ellis and Tavakol (CET) is based on an effective energy momentum tensor formed by the algebraic decomposition of the 4th order Bel-Robinson tensor. So far the application of the CET proposal has been limited to  spacetimes of Petrov types D and N for which this algebraic decomposition is unique. To address this limitation we examine in detail the effective energy momentum tensors that result from the algebraic decomposition of the Bel-Robinson tensor in Petrov type I spacetimes. As a test case we apply these results to the Szekeres models of class II, a Petrov type I analytic solution.  
\end{abstract}

\section{Introduction}\label{intro}
Observational evidence and theoretical considerations (the near homogeneity and isotropy of the CMB) \cite{aghanim2020planck, arbey2021dark,grohs2023big,Bennett_1996,Miller_1999,Lee_2001,Spergel_2003,Bennett_2003,Hinshaw_2009} strongly suggest the early universe up to matter-radiation decoupling to be close to thermal equilibrium. However, an entropy defined in terms of thermal interactions should necessarily reach a maximum and stop growing when structure formation sets in and the Universe becomes gravity dominated and non-thermal \cite{penrose1989difficulties, ellis2002cosmology}. 

Given this outlook, Penrose \cite{penrose1979singularities} conjectured that some notion of gravitational entropy was needed to account for structure formation that takes place when thermal processes that increased entropy in the early Universe plasma are no longer dominant. Penrose also remarked that the Weyl tensor can be thought of as encoding the free gravitational field, since it is nonzero even in the absence of sources, in which case the Ricci tensor vanishes (hence it is the curvature associated with the sources). 

Bearing in mind these theoretical connections, Penrose proposed the concept of a gravitational ``arrow of time'' that guides self-gravitating systems undergoing structure formation to evolve through timelike directions in which an early dominant Ricci curvature should be overtaken by an increasing Weyl curvature as the Universe evolved away from the early plasma into a non-linear structure formation process associated with the long range gravitational interaction. 

Penrose and other authors \cite{penrose1979singularities, wainwright1984, goode1985} have speculated that when this ratio is non-decreasing it should signal a direction of increasing inhomogeneity, producing an increase of entropy in the context of self-gravitating systems in the framework of the microcanonical ensemble. As the system becomes more inhomogeneous a wider range of densities and momenta are accessed, which increases the volume of accessible states in the phase space. In more recent appraisals of Penrose's proposal \cite{pelavas2006gravitational, zhao2018black, guha2020gravitational, gregoris2022understanding} the Ricci scalar can be replaced by the Kretschmann scalar that is nonzero even in vacuum solutions (allowing for the application to vacuum spacetimes). 

Considering the role of Weyl curvature in conceiving a gravitational entropy, Clifton, Ellis and Tavakol (CET) proposed \cite{clifton2013gravitational} a different gravitational entropy that also regards Weyl curvature as encoding the free gravitational field, but instead of proposing a scalar formulation, their proposal is based on an effective gravitational energy-momentum tensor constructed from the algebraic decomposition of the Bel-Robinson tensor, which is the only divergence-free tensor that can be constructed from the Weyl tensor. By introducing an effective energy-momentum tensor and a Gibbs one-form, the CET entropy proposal is  theoretically more robust than previous formulation by Penrose and others based on curvature scalars.

However, the Bel-Robinson tensor is fourth order, hence CET considered as a candidate for the effective energy-momentum tensor the second order tensors that result from the algebraic decomposition of the Bel-Robinson tensor as a tensor product of two second order tensors (i.e. the ``square root'' of the Bel-Robinson tensor). This algebraic decomposition of the Bel-Robinson tensor had been examined by Bonilla and Senovilla \cite{bonilla1997some, bonilla1997very, bonilla1998miscellaneous}, showing that it is only unique and non-degenerate for Petrov types N and D (wavelike and Coulomb-like spacetimes), with a degeneracy of non-unique second order tensors for the remaining Petrov types. 

The unique second order root of the Bel-Robinson tensor for spacetimes of Petrov types D and N, can be invariantly projected in terms of an arbitrary 4-velocity field and its orthogonal rest spaces, leading to the definition of ``gravitational" state variables: density, energy flux and stress tensor, with these variables being unrelated to the state variables associated with the matter energy-momentum tensor \cite{pelavas2006gravitational, pelavas2000measures}. From these geometric state variables, CET construct the 4-velocity projected component of the thermodynamical Gibbs one form, that defines a convective derivative of a gravitational entropy, with a gravitational temperature as its integrating factor. Since this procedure does not determine the gravitational temperature (as would be the case with an equation of state in a thermal system), CET introduce a gravitational temperature in an ad hoc manner in terms of the definition of gravitational redshift between local comoving observers. 

Once the geometric state variables have been determined for a given spacetime (i.e. a solution of Einstein's equations), the next task is to verify if the gravitational entropy increases with proper time. This task has been accomplished for LTB \cite{sussman2015gravitational} and Szekeres class I dust models \cite{pizana2022gravitational}, showing that CET entropy grows in inhomogeneities (over-densities and voids) that can be described by the exact generalization of the growing mode of dust perturbations. In particular, \cite{sussman2015gravitational, pizana2022gravitational} show that a positive cosmological constant provides a finite asymptotic saturation value for the CET gravitational entropy. However, other references that have examined the CET entropy  \cite{gregoris2020thermodynamics} show that the magnitude of the ratio of Weyl to Ricci curvature decreases with cosmic expansion, a result that contradicts the assumptions of the CET proposal. Furthermore, in \cite{chakraborty2022appropriate} CET entropy has also been applied to wormholes and compared to other gravitational entropy definitions, providing a unique gravitational entropy for Petrov D and N spacetimes. 

In this article we consider the case of Petrov type I spacetimes,  for which there is no unique algebraic decomposition of the Bel-Robinson tensor as a tensor product of two second order tensors (i.e. the ``square root'' of the Bel-Robinson tensor has several factors). Following Bonilla and Senovilla \cite{bonilla1997some, bonilla1997very, bonilla1998miscellaneous}, we examine the second order tensors whose tensor product reproduces the Bel-Robinson tensor, corresponding to the degeneracy of non-unique second order tensors for Petrov type I.  As a test case application, we focus on an exact Petrov type I solution: the Szekeres class II models whose source admits a nonzero energy flux and that has potentially interesting  cosmological applications.   We also  examine the integrability of the Gibbs one-form in connection with Einstein's equation in the 1+3 fluid flow approach, showing that the gravitational variables of the CET entropy can be related to the 4-acceleration, energy flux and magnetic Weyl tensor  that are the main drivers of inhomogeneity of the models.

The section by section content of the paper is described as follows. Section \ref{sec:CET} summarizes the CET gravitational entropy proposal. In Section \ref{sec:PeTyI}, we examine in detail the effective energy momentum tensors resulting from the algebraic decomposition of the Bel-Robinson tensor for Petrov Type I spacetimes. As an application of the results of Section \ref{sec:PeTyI}, we examine in Section  \ref{SzekCET} the case of Szekeres models of class II, an analytic solution of Petrov Type I. Our conclusions are stated in Section \ref{concl}. We discuss in \ref{sec:divfree-tracefree entropy relation} the relation between a traceless effective energy-momentum tensor and an effective energy-momentum tensor satisfying energy conservation. Lastly, in \ref{AppB}, we present some additional results associated to the gravitational states of the factors of the Bel-Robinson tensor in Petrov type I spacetime. 

\section{The gravitational entropy proposal of Clifton, Ellis and Tavakol (CET)}\label{sec:CET}

The CET entropy formalism focuses of the Bel-Robinson tensor, the unique divergence-free tensor that can be constructed solely from the Weyl tensor:
\begin{equation} \tensor{{\cal T}}{_a_b_c_d}=\tensor{C}{_e_a_b_f}\tensor{C}{^e_c_d^f}+\tensor{C}{^*_e_a_b_f}\tensor{C}{^*^e_c_d^f},\label{BelRob}\end{equation}
where \footnote{We drop the factor of $\frac{1}{4}$ from the definition of the Bel-Robinson tensor following the convention of \cite{bonilla1997some}. If one wishes to follow the convention of \cite{clifton2013gravitational} using this factor, it can be absorbed in the coupling constants of the ``gravitational'' state variables.} $C^*_{abcd}=\frac12 \eta_{abef}C^{ef}_{\;\;\;\;cd}$ is the dual Weyl tensor and $\eta_{abcd}$ is the completely anti-symmetric Levi-Civita tensor with $\eta_{0123}=-\sqrt{-\textrm{det}(g)}$. A symmetric, traceless second order tensor $t_{ab}$ can, in general, be derived from a fourth order symmetric divergence-free tensor ${\cal F}_{abcd}$, through an algebraic decomposition commonly referred to as its ``square root'' \cite{bonilla1997some, bonilla1997very, bonilla1998miscellaneous}.
\begin{equation} \tensor{{\cal F}}{_a_b_c_d}=\tensor{t}{_(_a_b}\tensor{t}{_c_d_)}-\frac12 \tensor{t}{_e_(_a}\tensor{t}{_b^e}\tensor{g}{_c_d_)}-\frac14 \tensor{t}{^e_e}\tensor{t}{_(_a_b} \tensor{g}{_c_d_)}+\frac{1}{24}\left(\tensor{t}{_e_f}\tensor{t}{^e^f}+\frac12 (\tensor{t}{^e_e})^2\right)\tensor{g}{_(_a_b}\tensor{g}{_c_d_)}.\label{decomp}\end{equation}
An interesting property of the square root tensor is that, for a particular solution $t_{ab}$ of Eq.~(\ref{decomp}),
\begin{eqnarray}
    {\cal T}_{ab}&=\epsilon t_{ab}+fg_{ab}
\end{eqnarray}
is the most general solution where $f$ is an arbitrary scalar function and $\epsilon^2=1$ \cite{bonilla1997some}. 

In the case of the Bel-Robinson tensor, CET interprets this second order symmetric tensor ${\cal T}_{ab}$ as an ``effective'' energy-momentum tensor associated with the Weyl curvature ({\it i.e.} the free gravitational field). It is important to emphasize that ${\cal T}_{ab}$ does not correspond to a physical source term to be included on the right hand side of Einstein's equations, but rather constitutes a formal, geometrically defined energy-momentum tensor. Nevertheless, given a 4-velocity field $u^{a}$ and the associated projection onto its orthogonal rest space, $h^{ab}=u^a\,u^b+g^{ab}$, this energy-momentum tensor allows for the definition of ``gravitational state variables'' ({\it i.e.} gravitational density, pressure, anisotropic pressure and energy flux) .
\begin{equation}{\cal T}^{ab}=\rho_{_\textrm{\tiny{gr}}}u^{a}\,u^{b}+p_{_\textrm{\tiny{gr}}}h^{ab}+\Pi_{_\textrm{\tiny{gr}}}^{ab}+2 q_{_\textrm{\tiny{gr}}}^{(a}u^{b)}\,.\label{Tabgr}\end{equation}
To fix the arbitrary function $f$, CET assumes the condition of energy conservation ($u_a\nabla_b\mathcal{T}^{ab}=0$) in vacuum.

Motivated by the Gibbs relation $T{\bf d} S = {\bf d} {\cal E}+ p {\bf d} V$, where ${\cal E}=\rho V$ represents the energy contained in a local volume $V$ (where the volume expansion satisfies $\dot V/V=\Theta=\nabla_au^a$) and $T$ is the temperature, CET extends this analogy in the gravitational sector associated to the Weyl curvature as \cite{clifton2013gravitational}
\begin{eqnarray}
    \Tgr \dd s &\equiv   \dd(\rho_{_\textrm{\tiny gr}} \V ) + p_{_\textrm{\tiny{gr}}} \dd \V\\
           &= \dd \rho_{_\textrm{\tiny gr}}\, \mathcal{V} + (\rho_{_\textrm{\tiny gr}}+p_{_\textrm{\tiny{gr}}})\dd{\mathcal{V}}\label{Tds1}.
\end{eqnarray}
Projecting the Gibbs one form of the gravitational variables on the 4-velocity field $u^a$, the gravitational entropy growth is given by
\begin{eqnarray}\label{sdot-gr}
    \Tgr \dot s &= \dot{\rho}_{_\textrm{\tiny gr}} \mathcal{V} + (\rho_{_\textrm{\tiny gr}}+ p_{_\textrm{\tiny{gr}}})\dot{\mathcal{V}}\label{Tds2}.
\end{eqnarray}
The analogy with the Gibbs one-form does not determine the gravitational temperature $T_{_\textrm{\tiny{gr}}}$. Consequently, CET introduces it through the local redshift experienced by comoving observers, expressing it in terms of the kinematical  parameters arising from the decomposition of $\nabla_b u_a$
\begin{equation}  T_{_\textrm{\tiny{gr}}} =\frac{\left| \nabla_a u_b\, k^{a}l^{b}\right|}{\pi},\label{Tgrav}\end{equation}
where $k^{a}\,,\,l^{a}$ denote the real null vectors associated with the Newman Penrose (NP) tetrad and $ \nabla_a u_b$ is expressed, in a coordinate basis, as
 \begin{equation} \nabla_a u_b = \frac13\Theta h_{ab}-\dot u_a u_b +\sigma_{ab}+\omega_{ab},\label{GradU}\end{equation}
where one identifies the expansion scalar $\Theta=\nabla_a u^a$, the 4-acceleration $\dot u_a = u^b \nabla_b u_a$, the shear tensor $\sigma_{ab}=\nabla_{(a} u_{b)}-\frac13 \Theta h_{ab}$ and the vorticity tensor $\omega_{ab}=\nabla_{[a} u_{b]}$. 

Because $T_{_\textrm{\tiny{gr}}}$ is non-negative by construction, the condition for gravitational entropy production becomes both necessary and sufficient when written as
\begin{eqnarray}
\dot{\rho}_{_\textrm{\tiny gr}} \mathcal{V} + (\rho_{_\textrm{\tiny gr}}+ p_{_\textrm{\tiny{gr}}})\dot{\mathcal{V}}&\geq 0.\label{entprod0}\end{eqnarray}

\section{Extension of CET entropy to Petrov type I spacetimes}\label{sec:PeTyI}

While any symmetric, divergence-free $t_{ab}$ uniquely determines a fourth order symmetric divergence-free tensor ${\cal F}_{abcd}$ via Eq.~(\ref{decomp}), the converse does not hold. In general, a given ${\cal F}_{abcd}$ does not correspond to a unique $t_{ab}$. In particular, for the Bel-Robinson tensor Eq.~(\ref{BelRob}), a unique second order symmetric divergence-free tensor arises  only in spacetimes of Petrov type D and N. Since the model under consideration is Petrov type I, we will use the ``factorization'' derived by Bonilla and Senovilla for such spacetimes \cite{bonilla1997some}.

\textbf{Definition:} \textit{Any completely symmetric traceless tensor $\mathcal F_{abcd}$ can be factorized into the traceless symmetric factors $A_{ab}$ and $B_{ab}$ if it can be written in the form}
\begin{eqnarray}
    \mathcal F_{abcd} &= \textrm{PS}[A_{ab}B_{cd}]
\end{eqnarray}
with $A^a_{\,a}=B^a_{\,a}=0$ and with $\textrm{PS}[A_{ab}B_{cd}]$ defined as
\begin{eqnarray}
\mathrm{PS}[A_{ab}B_{cd}] &\equiv&A_{(ab} B_{cd)}-\frac{1}{2} A_{\sigma(a} B_{b}^{\;\;\sigma} \, g_{cd)}-\frac{1}{8}\left(
A^{\rho}{}_{\rho} B_{(ab} g_{cd)}+
B^{\rho}{}_{\rho} A_{(ab} g_{cd)}
\right)\nonumber\\
&&+\frac{1}{24}\left(
A_{\rho\sigma} B^{\rho\sigma}
+\frac{1}{2} A^{\rho}{}_{\rho} B^{\sigma}{}_{\sigma}\right) g_{ab} g_{cd}.
\end{eqnarray}
As a corollary, we can have the most general factorization of any completely symmetric traceless tensor $\mathcal F_{abcd}$ as 
\begin{eqnarray}\label{trace-term-corollary}
    \mathcal F_{abcd} &= \textrm{PS}[\tilde A_{ab}\tilde B_{cd}],
\end{eqnarray}
such that the factors $\tilde A_{ab}$ and $\tilde B_{cd}$ are related to the traceless factors $\tilde A_{ab}$ and $\tilde B_{cd}$ by
\begin{eqnarray}
    \tilde A_{ab} &= \frac{1}{\mathcal{G}}\left(A_{ab}+ h_A\, g_{ab}\right),\\
    \tilde B_{ab} &= \mathcal{G}\left(B_{ab}+ h_B\, g_{ab}\right),
\end{eqnarray}
where $h_A,h_B$ and $\mathcal{G}$ are three arbitrary functions with $\mathcal{G}\neq 0$. $\tilde A_{ab}$ and $\tilde B_{cd}$ are no longer traceless. Going from a traceless effective energy momentum to a divergence-free energy momentum changes the gravitational density $\rho_{_\textrm{\tiny gr}}$ and the gravitational pressure $p_{_\textrm{\tiny gr}}$. All the other gravitational states remain unaffected (see \ref{sec:divfree-tracefree entropy relation}).

In order to construct the factors of the Bel-Robinson tensor Eq.~(\ref{BelRob}), we use the Weyl scalars as they contain the physical degrees of freedom of the Weyl tensor of the spacetime. For this purpose, we use the Newman-Penrose tetrad $\{\vec l,\vec k,\vec m,\vec{\bar m} \}$ and write the Weyl scalars as
\begin{eqnarray}\label{conformal-invars}
    \Psi_0 &= C_{abcd}l^am^bl^cm^d,\qquad \Psi_1 = C_{abcd}l^a k^bl^cm^d,\qquad \Psi_2 = C_{abcd}l^am^b\bar m^ck^d,\nonumber\\
   \Psi_3 &= C_{abcd}l^ak^b\bar m^c k^d,\qquad \Psi_4 = C_{abcd} k^a \bar m^b k^c \bar m^d.
\end{eqnarray}
For Petrov type I spacetimes, we can align $\vec l$ and $\vec k$ into any two of the four principal null directions which would give us $\Psi_0=\Psi_4=0$ \cite{Stephani:2003tm}. There are three possible factorizations of the Bel-Robinson tensor, which are given by
\begin{eqnarray}
    \mathcal T_{abcd} &= \textrm{PS}[A^1_{ab}\cdot B^1_{cd}] = \textrm{PS}[A^\pm_{ab}\cdot B^\pm_{cd}],
\end{eqnarray}
where each of the factors is symmetric traceless, and can be written in terms of the Weyl scalars in the NP basis as \cite{bonilla1997some}
\begin{eqnarray}
A^1_{ab}&= 6 |\Psi_2|\left[m_{(a}\bar m_{b)} + l_{(a} k_{b)} \right],\label{TT1}\\
B^1_{ab}&= 6 |\Psi_2|\Bigg[\frac{8|\Psi_3|^2}{9|\Psi_2|^2}l_al_b -\frac{4\Psi_3}{3\Psi_2}l_{(a}m_{b)} -\frac{4\bar\Psi_3}{3\bar\Psi_2}l_{(a}\bar m_{b)} +\frac{8|\Psi_1|^2}{9|\Psi_2|^2}k_a k_b -\frac{4\bar\Psi_1}{3\bar\Psi_2}k_{(a}m_{b)} \nonumber\\
    &\quad -\frac{4\Psi_1}{3\Psi_2}k_{(a}\bar m_{b)} + \frac{8\bar\Psi_1\Psi_3}{9|\Psi_2|^2}m_a m_b +\frac{8\Psi_1\bar\Psi_3}{9|\Psi_2|^2}\bar m_a \bar m_b +m_{(a}\bar m_{b)}+l_{(a}k_{b)}\Bigg], \label{TT2}\\
A^\pm_{ab}&= \pm\bigg[2(3\Psi_2\pm\Xi)l_al_b +\frac{8|\Psi_1|^2}{3\bar\Psi_2\pm\bar\Xi}\left(m_{(a}\bar m_{b)}+l_{(a}k_{b)}\right)-8\Psi_1 l_{(a}\bar m_{b)}\nonumber\\
&\quad -8\bar\Psi_1\frac{(3\Psi_2\pm\Xi)}{(3\bar\Psi_2\pm\bar\Xi)} l_{(a}m_{b)} \bigg],\label{TT3}\\
B^\pm_{ab}&=\pm\bigg[2(3\bar\Psi_2\pm\bar\Xi)k_a k_b+ \frac{8|\Psi_3|^2}{3\Psi_2\pm\Xi}\left(m_{(a}\bar m_{b)}+l_{(a}k_{b)}\right) -8\bar\Psi_3 k_{(a}\bar m_{b)}\nonumber\\
&\quad -8\Psi_3\frac{(3\bar\Psi_2\pm\bar\Xi)}{(3\Psi_2\pm\Xi)}k_{(a}m_{b)}\bigg]\label{TT4}.
\end{eqnarray}
where $\Xi=\sqrt{9\Psi_2^2-16\Psi_1\Psi_3}$. As opposed to \cite{bonilla1997some}, here we have chosen $A^\pm$ and $B^\pm$ to possess the correct physical dimensions by multiplying $A^\pm$ and dividing $B^\pm$ of \cite{bonilla1997some} with the scalar $\mathcal{F}_\pm= \pm (3\Psi_2\pm\Xi)$. This choice is justified as these factors converge to the correct Petrov type D limits illustrated in Section \ref{sec:limit-typeD}.

\subsection{Gravitational states from the traceless factors of the Bel-Robinson tensor}\label{Grav-States}
From the effective energy momentum tensor $\T_{ab}$ in Eq.~(\ref{Tabgr}), we can determine the gravitational states as
\begin{eqnarray}
\rho_{_\textrm{\tiny gr}} &= \T_{ab}u^a u^b\,, \quad p_{_\textrm{\tiny gr}} = \frac{1}{3}h_{ab} \T^{ab}\,, \quad  q^{\textrm{\tiny gr}}_a = - \tensor{h}{_a^b}\tensor{\T}{_b^c}u_c\,, \nonumber\\
\Pi^{\textrm{\tiny gr}}_{ab} &= \left[\tensor{h}{_(_a^c}\tensor{h}{_b_)^d}-\frac13 h_{ab}h^{cd}\right]\T_{cd}\,.
\end{eqnarray}
In the following section, we consider the traceless factors Eqs.~(\ref{TT1}-\ref{TT4}) to be the effective energy momentum tensor $\mathcal{T}_{ab}$ and use them to construct the gravitational state variables.
\begin{itemize}
    \item Gravitational states of $A^\pm_{ab}$ and $B^\pm_{ab}$\;:
    \begin{eqnarray}
        \kappa_{_{\textrm{\tiny{A}}^\pm}} \rho_{_{\textrm{\tiny{A}}^\pm}} &= \frac{4|\Psi_1|^2}{\bar\Xi\pm3\bar\Psi_2} +\Xi\pm 3\Psi_2\,, \qquad p_{_{\textrm{\tiny{A}}^\pm}}= \frac{1}{3}\rho_{_{\textrm{\tiny{A}}^\pm}},\label{Apm-states}\\
        \kappa_{_{\textrm{\tiny{B}}^\pm}} \rho_{_{\textrm{\tiny{B}}^\pm}} &= \frac{4|\Psi_3|^2}{\Xi\pm3\Psi_2} +\bar\Xi\pm 3\bar\Psi_2\,, \qquad p_{_{\textrm{\tiny{B}}^\pm}}= \frac{1}{3}\rho_{_{\textrm{\tiny{B}}^\pm}}\label{Bpm-states}.
    \end{eqnarray}

    \item Gravitational states of $A^1_{ab}$ and $B^1_{ab}$\;:
    \begin{eqnarray}
        \kappa_{_{\textrm{\tiny{A}}^1}} \rho_{_{\textrm{\tiny{A}}^1}} &= 3|\Psi_2|\,, \qquad p_{_{\textrm{\tiny{A}}^1}}= \frac{1}{3}\rho_{_{\textrm{\tiny{A}}^1}},\label{A1-states}\\
        \kappa_{_{\textrm{\tiny{B}}^1}} \rho_{_{\textrm{\tiny{B}}^1}} &= \frac{1}{3|\Psi_2|}\left(8|\Psi_1|^2+8|\Psi_3|^2+9|\Psi_2|^2\right) \,, \qquad p_{_{\textrm{\tiny{B}}^1}}= \frac{1}{3}\rho_{_{\textrm{\tiny{B}}^1}}\label{B1-states}.
    \end{eqnarray}
\end{itemize}
For brevity, we remove ``\textrm{gr}" subscript (or superscript) when talking about the gravitational state of a specific factor and instead replace it with the corresponding factor.
The expressions of the gravitational states $q^{\textrm{\tiny gr}}_a$ and $\Pi^{\textrm{\tiny gr}}_{ab}$ corresponding to $A^1_{ab},\,B^1_{ab}$  and $A^\pm_{ab},\,B^\pm_{ab}$ are put in \ref{AppB}, since they are not necessary to compute the gravitational entropy.
\subsection{Limiting case of Petrov type D}\label{sec:limit-typeD}
The Weyl tensor of Petrov type I is reduced to type D when $|\Psi_1|=|\Psi_3|\to 0$. In this regard, we define the ratio
\begin{eqnarray}
     \alpha &\equiv \frac{4}{3}\frac{|\Psi_1|}{|\Psi_2|} \left(\equiv \frac{4}{3}\frac{|\Psi_3|}{|\Psi_2|}\right) 
\end{eqnarray}
which determines the transition from a Petrov type I spacetime to a Petrov type D. For type D, the only non vanishing conformal invariant is $\Psi_2$ which determines the electric Weyl tensor. Eq.~(\ref{TT1}) and Eq.~(\ref{TT2}) reduce to the unique traceless square root $t_{ab}$ of the Bel-Robinson tensor for Petrov type D and is given by
\begin{eqnarray}
t_{ab} &= A^1_{ab}\big|_{\alpha=0} = B^1_{ab}\big|_{\alpha=0} = 6 |\Psi_2|\left[m_{(a}\bar m_{b)} + l_{(a} k_{b)} \right].
\end{eqnarray}
As for the factors $A^\pm$ and $B^\pm$ from Eq.~(\ref{TT3}) and Eq.~(\ref{TT4}), we need to carefully consider the following cases based on $\mathcal F_\pm=0$ which is also present in the denominator. $\mathcal F_\pm$ can be written as 
\begin{eqnarray}
\mathcal F_\pm &= 3|\Psi_2|\left(\textrm{sgn}(\Psi_2) \pm \sqrt{1-\alpha^2}\right)\\
&= 3|\Psi_2|\left(\mathrm{sgn}(\Psi_2) \pm 1\right) \quad\textrm{for $\alpha=0$.}
\end{eqnarray}
If $\textrm{sgn}(\Psi_2)=1$, $\mathcal{F}_- = 0$, which suggests the divergence of the factors $A^-_{ab}$ and $B^-_{ab}$. From Eqs.~(\ref{Apm-states}-\ref{Bpm-states}), it would also result in $\rho_{_{\textrm{\tiny{A}}^-}}<0$ and $\rho_{_{\textrm{\tiny{B}}^-}}<0$. This makes the factorization of the Bel-Robinson becomes invalid. However, as $\mathcal{F}_+ \neq 0$, the factorization of the Bel-Robinson into $A^+_{ab}$ and $B^+_{ab}$ is still valid with $\rho_{_{\textrm{\tiny{A}}^+}}>0$ and $\rho_{_{\textrm{\tiny{B}}^+}}>0$. For $\textrm{sgn}(\Psi_2)=-1$, $\mathcal{F}_+ = 0$ and the situation reverses and $A^+_{ab}$ and $B^+_{ab}$ diverge with $\rho_{_{\textrm{\tiny{A}}^+}}<0$ and $\rho_{_{\textrm{\tiny{B}}^+}}<0$ while $A^-_{ab}$ and $B^-_{ab}$ remain finite with $\rho_{_{\textrm{\tiny{A}}^-}}>0$ and $\rho_{_{\textrm{\tiny{B}}^-}}>0$. Thus, the factorization of the Bel-Robinson tensor into $A^\pm_{ab}\cdot B^\pm_{cd}$ depends on the $\textrm{sgn}(\Psi_2)$. In the case of $\alpha =0$, Eq.~(\ref{TT3}) and Eq.~(\ref{TT4}) recover the two pure radiation factors of the Bel-Robinson tensor for Petrov type D given by
\begin{eqnarray}
    A^\pm_{ab} &= 12 |\Psi_2|l_a l_b\quad \textrm{or}\quad 12 |\Psi_2|k_a k_b.
\end{eqnarray}

\iffalse
 expressed in an orthonormal tetrad $\{x_{_\textrm{\tiny{A}}},\,y_{_\textrm{\tiny{A}}},\,z_{_\textrm{\tiny{A}}},\,u_{_\textrm{\tiny{A}}}\}$ with $\textrm{\tiny{A}}=0,1,2,3$:
%
\begin{equation}8\pi {\cal T}_{_{_\textrm{\tiny{AB}}}}=\epsilon\alpha |\Psi_2|\left[x_{_\textrm{\tiny{A}}}x_{_\textrm{\tiny{B}}} +y_{_\textrm{\tiny{A}}}y_{_\textrm{\tiny{B}}} -2(z_{_\textrm{\tiny{A}}}z_{_\textrm{\tiny{B}}} -u_{_\textrm{\tiny{A}}} u_{_\textrm{\tiny{A}}})\right],\label{TT} \end{equation}
%
where $\epsilon=\pm 1$,\,\, $\alpha$ is a constant (to set units) and $\Psi_2= C_{_{_\textrm{\tiny{ABCD}}}}\,k^{_\textrm{\tiny{A}}}\,m^{_\textrm{\tiny{B}}}\,\tilde m^{_\textrm{\tiny{C}}}\,l^{_\textrm{\tiny{D}}}$ is the only nonzero Weyl scalar for Petrov type D spacetimes given in terms of the null tetrad associated with $\{x_{_\textrm{\tiny{A}}},\,y_{_\textrm{\tiny{A}}},\,z_{_\textrm{\tiny{A}}},\,u_{_\textrm{\tiny{A}}}\}$.
\fi

In the following section, we focus on a particular subclass of Petrov Type~I, \textit{viz.} the Szekeres Class II solutions characterized by the conditions $\Psi_1 =\bar \Psi_3$ and $\Psi_2,\Xi \in \mathbb{R}$. The condition $\Xi \in \mathbb{R}$ is justified as a consequence of the dominant energy condition (DEC). Moreover, we want these gravitational states to reproduce those of the Petrov type D for $\alpha\to 0$. As such, we change the coupling constants for $A^\pm$ and $B^\pm$ to $\kr$ for the pure radiation case of type D and $A^1$ and $B^1$ to $\kc$ for the Coulombic case. For $A^\pm_{ab}$ and $B^\pm_{ab}$, we get the same density and pressure given by
\begin{eqnarray}
   \kappa_{\textrm{\tiny{R}}} \rho_{\textrm{\tiny{R}}} &= \frac{15}{4}|\Psi_2|\left(1 +\frac{3}{5}\sqrt{1-\alpha^2}\right).
\end{eqnarray}
For $A^1_{ab}$ and for $B^1_{ab}$, they are
\begin{eqnarray}
\kappa_{\textrm{\tiny{C}}}\rho_{\textrm{\tiny{A}}_1}&= 3 |\Psi_2|,\quad\textrm{and}\quad \kappa_{\textrm{\tiny{C}}}\rho_{\textrm{\tiny{B}}_1}= 3|\Psi_2|\left(1+\alpha^2 \right).
\end{eqnarray} 
An interesting result to look into is that assuming $\kappa_{\textrm{\tiny{C}}}=\kappa_{\textrm{\tiny{R}}}$ at zero order in $\alpha$, we find that $\rho_{\textrm{\tiny{R}}}>\rho_{\textrm{\tiny{A}}_1}$ as well as $\rho_{\textrm{\tiny{R}}}>\rho_{\textrm{\tiny{B}}_1}$.
\subsection{Gravitational entropy growth associated to the traceless factors}
Now that the gravitational states are determined corresponding to each of the traceless factors of the Bel-Robinson tensor in the Petrov type I spacetime, we are in a position to calculate the rate of entropy production associated to them. Each factors has the equation of state $p_{\textrm{\tiny{gr}}}= \frac{1}{3}\rho_{_\textrm{\tiny{gr}}}$. Using this together with $\dot\V/\V=\Theta$ in Eq.~(\ref{sdot-gr}), we find that
\begin{eqnarray}
    \dot s&= \frac{\V}{\Tgr} \left(\dot{\rho}_{_\textrm{\tiny{gr}}} +\frac{4}{3}\Theta\rho_{\textrm{\tiny gr}}\right).
\end{eqnarray}
As the factors $A^\pm_{ab}$ and $B^\pm_{ab}$ have the same gravitational density $\rho^{\textrm{\tiny gr}}_{_\textrm{\tiny R}}$, so they would have the same entropy growth given by
\begin{eqnarray}\label{sdot-r}
    \dot{s}_{_{\tiny\textrm{R}}}&= \frac{15|\Psi_2|\V}{4\kappa_{_\textrm{\tiny R}}\Tgr}\Bigg[\frac{1}{|\Psi_2|}\pdv{|\Psi_2|}{t}\left(1+\frac{3}{5}\sqrt{1-\alpha^2}\right)+\frac{4}{3}\Theta\left(1+\frac{3}{5}\sqrt{1-\alpha^2}\right)\nonumber\\
    &\quad -\frac{3}{5}\frac{\alpha\dot\alpha}{\sqrt{1-\alpha^2}}\Bigg].
\end{eqnarray}
For $A^1_{ab}$ and for $B^1_{ab}$, the entropy growth is given by
\begin{eqnarray}
    \dot{s}_{_{\textrm{\tiny A}_1}}&= \frac{3|\Psi_2|\V}{\kappa_{\textrm{\tiny{C}}}\Tgr}\left[\frac{1}{|\Psi_2|}\pdv{|\Psi_2|}{t} +\frac{4}{3}\Theta \right],\label{sdot-A1}\\
    \dot{s}_{_{\textrm{\tiny B}_1}}&= \frac{3|\Psi_2|\V}{\kappa_{\textrm{\tiny{C}}}\Tgr}\left[\frac{1}{|\Psi_2|}\pdv{|\Psi_2|}{t}\left(1+\alpha^2\right) +2\alpha\dot\alpha+\frac{4}{3}\Theta\left(1+\alpha^2\right) \right].\label{sdot-B1}
\end{eqnarray}
We would like to point out an important issue here. Since a trace term can always be added to the factors of the Bel Robinson tensor (see Eq.~(\ref{trace-term-corollary})), so the gravitational states ($\rho_{\textrm{\tiny{gr}}}$ and $p_{\textrm{\tiny{gr}}}$) are not unique (see \ref{sec:divfree-tracefree entropy relation} for the effect of introducing a trace term in the effective energy–momentum tensor on the resulting gravitational states). As such, it also changes the gravitational entropy associated to the factors. One way, prescribed by CET, is to fix the trace term such that the effective energy momentum satisfies energy conservation in vacuum. However, for the purpose of this paper, we look at the gravitational entropy associated only with the traceless factors as they exhibit some interesting features. 
\subsection{Integrability criteria of the Gibbs one-form for gravitational entropy}\label{sec:integrabilitycriteria}
To evaluate the gravitational entropy $s$ we need to integrate the Gibbs form along $u^a$: 
\begin{eqnarray}   s&= \int \frac{\dd\tau}{\Tgr}\left(\dot\rho_{_\textrm{\tiny gr}} \mathcal{V} + (\rho_{_\textrm{\tiny gr}}+p_{_\textrm{\tiny gr}})\dot{\mathcal{V}}\right)\label{entprod2}.\end{eqnarray}
The strongest notion of integrability is exactness. Consider a one-form $\omega$ whose components $\omega_a$ are expressed in a local coordinate system $\{x^a\}$ given by 
\begin{eqnarray}
    \omega &= \omega_a \dd x^a.
\end{eqnarray}
The one-form is said to be exact if there exists a scalar $\phi$ such that
\begin{eqnarray}
    \omega_a &= \nabla_a \phi.
\end{eqnarray}
In that case, $\omega_a$ is the gradient of a potential and the line integral of $\omega_a$ is path independent. A necessary and locally sufficient condition for exactness is that $\omega_a$ be closed, \textit{i.e.}
\begin{eqnarray}
    \nabla_{[a}\omega_{b]}&=0.
\end{eqnarray}
As the Levi-Civita connection is torsion-free, this can also be written as 
\begin{eqnarray}
    \partial_{[a}\omega_{b]}&=0.\label{int-cond}
\end{eqnarray}
Therefore, if we take the entropy one-form $\omega=\dd s$ with components being $\omega_a=\partial_a s$, then using Eq.~(\ref{int-cond}) we get the integrability condition for the entropy as
\begin{eqnarray}
    \partial_a\partial_b s -  \partial_b\partial_a s&=0.\label{int-cond2}
\end{eqnarray}
Using Eq.~(\ref{Tds1}) yields
\begin{eqnarray}
    \partial_a\left[\frac{1}{\Tgr}\left(V\partial_b\rho_{_\textrm{\tiny gr}} + (\rho_{_\textrm{\tiny gr}}+ p_{_\textrm{\tiny gr}})\partial_b\V\right)\right]-\partial_b\left[\frac{1}{\Tgr}\left(V\partial_a\rho_{_\textrm{\tiny gr}} + (\rho_{_\textrm{\tiny gr}}+p_{_\textrm{\tiny gr}})\partial_a\V\right)\right]&=0,\nonumber\\
\end{eqnarray}
which, along with the equation of state $\rho_{_\textrm{\tiny gr}}=3 p_{_\textrm{\tiny gr}}$(which holds for all the traceless factors) simplifies to
\begin{eqnarray}
    \Tgr \partial_{[a}p_{_\textrm{\tiny gr}}\partial_{b]}\V+3\V\partial_{[a}p_{_\textrm{\tiny gr}}\partial_{b]}\Tgr +4p_{_\textrm{\tiny gr}}\partial_{[a}\V\partial_{b]}\Tgr&=0.
\end{eqnarray}

\section{Szekeres-II models and the CET gravitational entropy}\label{SzekCET}

We now focus on a particular class of Petrov I spacetimes, Szekeres class II models \cite{najera2021pancakes}. The line element of Szekeres class II models in the comoving synchronous gauge is written as
\begin{equation}\label{eq:szekmet}
\dd s^2 = -\dd t^2 +S^2(t)\left[X^2(t,w,x,y)\dd w^2+\frac{\dd x^2+\dd y^2}{f^2(x,y)} \right],
\end{equation}
where
\begin{equation}
    f(x,y)=1+\frac{k\left(x^2+y^2\right)}{4}\,,\quad\textrm{and}\quad k=0,\pm 1.
\end{equation}
The orthonormal tetrad associated with this metric is given by
\begin{eqnarray}
    t_a = -\dd t\, \tensor{\delta}{^t_a},\quad w_a=SX\dd w \,\tensor{\delta}{^w_a},\quad x_a=\frac{S}{f}\dd x\,\tensor{\delta}{^x_a}, \quad y_a=\frac{S}{f}\dd y\, \tensor{\delta}{^y_a}. 
\end{eqnarray}
These solutions are compatible with the following energy-momentum tensor in a comoving frame
\begin{equation}
T^{ab} = \rho u^a u^b +ph^{ab}+\Pi^{ab}+2q^{(a}u^{b)},
\end{equation}
where the energy density $\rho$, isotropic pressure $p$, anisotropic stress $\Pi_{ab}=[\tensor{h}{_(_a^c}\tensor{h}{_b_)^d}-\frac13 h_{ab}h^{cd}]T_{cd}$, and energy flux $q_a= - \tensor{h}{_a^b}\tensor{T}{_b^c}u_c$ (with $h_{ab}=u_au_b+g_{ab}$) are, in general, functions of the four coordinates $t, x^i$.

The rest frames orthogonal to the comoving four-velocity $u_a = t_a$ are conformally flat (but not flat). The two-surfaces defined by constant $t$ and $w$ in Eq.~(\ref{eq:szekmet}) have constant curvature whose sign is determined by $k$, this leads to three subclasses: quasi-spherical ($k=1$), quasi-plane ($k=0$), and quasi-hyperbolic ($k=-1)$ models. In \cite{najera2021pancakes} it is proven that  only the quasi-plane models admit a smooth matching to a FLRW model. From here onwards, we restrict our analysis to the quasi-planar case.

The only non-vanishing kinematic quantities associated with the four-velocity $u^a$ are the expansion scalar $\Theta$ and the shear tensor $\sigma_{ab}$. In terms of the metric parameters, these quantities take the form
\begin{equation}
	\Theta = \frac{\dot X}{X}+3\frac{\dot S}{S}, \quad \tensor{\sigma}{^a_b}=\sigma \tensor{\xi}{^a _b}, \quad \sigma = -\frac{\dot X}{3X},
\end{equation}
where $\dot X=u^a  X_{,a}$ and $\tensor{\xi}{^a _b} = \tensor{\delta}{^a_w}\tensor{\delta}{^w_b}-3\tensor{h}{^a_b} = \textrm{diag}[0,-2,1,1].$
The Newman-Penrose (NP) null tetrad associated to this orthonormal basis can be constructed as
\begin{eqnarray}
    l_a&=\frac{1}{\sqrt{2}}\left( t_a + w_a\right), \qquad k_a=\frac{1}{\sqrt{2}}\left( t_a - w_a\right),\\
    m_a&=\frac{1}{\sqrt{2}}\left( x_a + i y_a\right), \qquad \bar m_a=\frac{1}{\sqrt{2}}\left( x_a - i y_a\right).
\end{eqnarray}
The Szekeres class II solutions with $q_a\neq 0$ are Petrov type I \cite{najera2021pancakes} while $q_a=0$ (whether Szekeres class I or II) are Petrov type D. For a simplified analysis of Petrov type I, we consider the energy-momentum tensor sourced by a matter density and an energy flux, \textit{i.e.} 
\begin{eqnarray}
    T^{ab}&= \rho u^a u^b + 2\tensor{q}{^(^a}\tensor{u}{^b^)}.    \label{EMEFE}
\end{eqnarray}
It was shown in \cite{najera2021non} that Szekeres class II spacetimes with such an energy-momentum tensor lead to cosmological models that are able to reproduce, in a qualitative manner, the magnitudes of observed peculiar velocities of galaxies in the Universe. Therefore,  our analysis is physically motivated, albeit simplified.

From Einstein's field equations we get
\begin{eqnarray}
    &\kappa \rho = \kappa \bar\rho -2\left(\frac{\ddot X}{X}+\frac{\dot S}{S}\frac{\dot X}{X} \right),\quad \kappa \bar\rho =3 \left(\frac{\dot S}{S}\right)^2,\label{EFE-1}\\
    &\kappa q_x = \frac{\dot X_{,x}}{X}, \quad \kappa q_y = \frac{\dot X_{,y}}{X},\label{EFE-2}\\
    &\dot S^2 +2S\ddot S =0,\\
    &X_{,xy}=0, \quad X_{,xx}= X_{,yy}= S\left(3\dot S\dot X+S\ddot X \right).
\end{eqnarray}
The non-vanishing Weyl scalars are
\begin{eqnarray}
    \Psi_1&=\bar \Psi_3 = -\frac{1}{4S}\left(\frac{\dot X_{,x}}{X} + i \frac{\dot X_{,y}}{X}\right),\label{CF-1} \quad\textrm{and}\\
    \Psi_2 &=-\left( \frac{2}{3}\frac{\dot S}{S}\frac{\dot X}{X}+\frac{1}{3}\frac{\ddot X}{X} \right)\label{CF-2},
\end{eqnarray}
which generate the magnetic ($H_{\langle ab\rangle}$) and electric ($E_{\langle ab\rangle}$) Weyl tensors, respectively
\begin{eqnarray}
\tensor{E}{_\langle_a_b_\rangle}&= \tensor{C}{_a_c_b_d}u^c u^d = \mathcal{E} \tensor{\xi}{_a_b}, \\
\tensor{H}{_\langle_a_b_\rangle}
&= \frac{1}{2}\tensor{\eta}{_a_c^e^f}\tensor{C}{_e_f_b_d} u^c u^d = \left(
\begin{array}{cccc}
0 & 0 & 0 & 0 \\
0 & 0 & H_{21} & H_{31} \\
0 & H_{21} & 0 & 0 \\
0 & H_{31} & 0 & 0
\end{array}
\right)_{ab},
\end{eqnarray}
where
\begin{eqnarray}
    \cal E &= \frac{2}{3}\frac{\dot S}{S}\frac{\dot X}{X}+\frac{1}{3}\frac{\ddot X}{X}\;,\qquad H_{21} =\frac{1}{2}S\dot X_{,y}\;,\qquad H_{31} =-\frac{1}{2}S\dot X_{,x}.
\end{eqnarray}
Rewriting in terms of the Weyl scalars, we find that,
\begin{eqnarray}
    \mathcal{E}=-\Psi_2, \quad H_{21}= i S^2 X (\Psi_1-\bar\Psi_1), \quad H_{31}=  S^2 X (\Psi_1+\bar\Psi_1).
\end{eqnarray}

\subsection{Coupling the CET gravitational entropy to the Einstein field equations}\label{CETGR}
It is evident that CET gravitational entropy  is closely linked to the dynamics of the conformal Weyl invariants $\Psi_1,\Psi_2$, and $\Psi_3$. In the 1+3 language, their evolution would be related to the physical sources (which in our case are the density $\rho$ and the energy flux $q_a$) through the electric and the magnetic Weyl tensors. For the Szekeres II case, the relation can be easily obtained by substituting the source from Eqs.~(\ref{EFE-1}, \ref{EFE-2}) in Eqs.~(\ref{CF-1}, \ref{CF-2}), which gives
\begin{eqnarray}
    \Psi_2 &= \frac{\kappa}{6}\left(\rho-\bar\rho \right),\\
    \Psi_1 &=\bar\Psi_3 = -\frac{\kappa}{4 S}\left(q_x+iq_y \right).
\end{eqnarray}
In terms of the sources, $\alpha$ can be written as
\begin{eqnarray}
\alpha&= \frac{2|q|}{|\rho-\bar\rho|},
\end{eqnarray}
where $|q|\equiv \sqrt{q^a q_a}$.
The 1+3 evolution equations of the sources for Szekeres II are
\begin{eqnarray}
    \dot\rho &= -\rho\Theta-\bar\nabla^a q_a,\\
    \tensor{\dot q}{_\langle_a_\rangle} &= -\frac{4}{3}\Theta q_a-\sigma_{ab}q^b.
\end{eqnarray}
Using these equations, we can determine the two quantities of interest that are related to the entropy growth, \textit{viz}, $\dot\Psi_2$ and $\dot\alpha$. They are given by
\begin{eqnarray}
\dot\Psi_2&= \frac{\kappa}{6}\left(-\rho\Theta-\bar\nabla^a q_a+\bar\rho\bar\Theta\right) ,\\
 \dot \alpha &= \alpha\left[-\frac{4}{3}\Theta -\frac{1}{|q|^2}\sigma_{ab}q^a q^b +\frac{\textrm{sgn}(\rho-\bar \rho)}{|\rho-\bar\rho|}\left(\rho\Theta +\bar\nabla^aq_a-\bar\rho\bar\Theta\right)\right].
\end{eqnarray}
Substituting them in Eqs.~(\ref{sdot-r}), (\ref{sdot-A1}) and (\ref{sdot-B1}), we get the expression of the entropy growth in terms of the sources and covariant scalars
\begin{eqnarray}
 \dot{s}_{_{\textrm{\tiny R}}}&= \frac{15|\Psi_2|\V}{4\kappa_{_\textrm{\tiny R}}\Tgr}\Bigg[\left(1+\frac{3}{5}\sqrt{1-\alpha^2}\right)\left(M+\frac{4}{3}\Theta\right) +\frac{3\alpha^2}{5\sqrt{1-\alpha^2}}\Bigg(\frac{4}{3}\Theta+N\nonumber\\
 &\quad-M\Bigg)\Bigg],\\
\dot{s}_{_{\textrm{\tiny A}_1}}&=   \frac{3|\Psi_2|\V}{\Tgr\kc}\left[\frac{4}{3}\Theta+M \right],\label{entropy-A1-sources}\\
   \dot{s}_{_{\textrm{\tiny B}_1}}&= \frac{3 |\Psi_2|\V}{\Tgr\kc}\left(1+\alpha^2\right)\left[\frac{4}{3}\Theta + M - \frac{2\alpha^3}{\left(1+\alpha^2\right)}\left(\frac{4}{3}\Theta +N +M\right)\right],\label{entropy-B1-sources}
\end{eqnarray}
where the variables $M$ and $N$ are defined as 
\begin{eqnarray}
    M &\equiv \frac{1}{(\rho-\bar\rho)}\left(-\rho\Theta -\bar\nabla^aq_a+\bar\rho\bar\Theta\right), \quad \textrm{and}\quad N\equiv\frac{1}{|q|^2}\sigma_{ab}q^a q^b.
\end{eqnarray}
To understand the physical implications of the entropy growth in the structures associated with Szekeres II solutions, we rewrite $\rho$ and $\Theta$ as the sum of an FLRW background part plus an exact fluctuation:
\begin{eqnarray}
    \rho&=\bar\rho+\mathcal{D}_\rho \quad\textrm{and}\quad \Theta = \bar\Theta+\mathcal{D}_\Theta.
\end{eqnarray}
Since the leading term associated to the entropy production in each of the cases is $L\equiv M+\frac{4}{3}\Theta$, we can expressed it in terms of the fluctuations $\mathcal{D}_\rho$ and $\mathcal{D}_\Theta$ that control the inhomogeneous deviation from an FLRW background: 
\begin{eqnarray}
    L &= \frac{1}{3}\Theta-\bar\rho\frac{\mathcal{D}_\Theta}{\mathcal{D}_\rho} -\bar\nabla^aq_a.
\end{eqnarray}
 This relation makes it evident that for the entropy to grow with the proper time of the fundamental observers, we should have at leading order $L\geq0$.
 
For an expanding universe characterized by $\Theta\geq0$, a sufficient condition for the entropy to grow is  $\mathcal{D}_\Theta/\mathcal{D}_\rho\leq0$ if the divergence of the energy flux is assumed to be small. Since this condition holds in a growing mode (see \cite{najera2021pancakes}), the gravitational entropy increases.  On the other hand,  growth of gravitational entropy is still possible  in a decaying mode as long as  $\Theta/3>\bar\rho \mathcal{D}_\Theta/\mathcal{D}_\rho +\bar\nabla^aq_a$ holds. The $\mathcal{O}(\alpha^2)$ correction follows the same pattern.

While, so far, we have discussed all the factors associated with type I spacetimes and their corresponding gravitational entropies, we propose that there is a natural choice for a unique effective energy-momentum tensor for a Szekeres II spacetime: the factor $\bf{B^1}$. This choice is motivated by the limiting cases of type D spacetimes. As shown in Section \ref{sec:limit-typeD}, while the factors $\bf{A^\pm}$ and $\bf{B^\pm}$ converge to the radiation factors, $\bf{A^1}$ and $\bf{B^1}$ are the only factors that converge to the unique square root in type D limit. More importantly, $\bf{A^1}$ retains the same form of the square root of the limiting type D in type I spacetimes and, as such, describes only the gravitational degrees of freedom associated with the electric Weyl tensor as in a type D spacetime. By constrast, the new information arising in the type I case, \textit{viz.}, the magnetic Weyl tensor, is contained in $\bf{B^1}$ in the Szekeres II spacetime.  This new information, compared to a type D density inhomogeneity, is the presence of an energy flux. It results in a non-vanishing magnetic Weyl tensor, which is contained only in $\bf{B^1}$. 

Following this line of reasoning, one might question about the presence of the divergence of the energy flux term (which clearly represents the new information of the type I spacetime) in the entropy growth associated with $\bf{A^1}$ in Eq. \eqref{entropy-A1-sources}. To address this, we note that, later in Section~\ref{CETGR}, we subject the growth of the gravitational entropy to the Einstein field equations. After this substitution, the resulting expression depends on the total content of the physical energy-momentum tensor. This is why the divergence of the energy flux term appears in Eq. \eqref{entropy-A1-sources}, its presence there does not imply that this information is intrinsically encoded in the factor $\bf{A^1}$.   

For the Szekeres II spacetime, the expression for the CET entropy growth (associated with the factor $\bf{B^1}$) in terms of the metric functions is given by
\begin{eqnarray}
    \dot s_{\textrm{\tiny CET}}&=& \frac{\mathcal{V}}{T_{\textrm{\tiny gr}}}\Bigg[-6\left(\frac{\dot S}{S}\right)^2\frac{\dot X}{X}-20\frac{\dot S}{S}\frac{\ddot X}{X}-4\frac{\dddot{X}}{X}+\frac{2}{3SX^2(2\dot S\dot X+S\ddot X)}\Big(-8\dot S^2\dot X^3 \nonumber\\
    &&+\dot X(\dot X_{,x})^2+\dot X(\dot X_{,y})^2-8S\dot S\dot X^2\ddot X - 2S^2\dot X\ddot X^2\Big)\Bigg],
\end{eqnarray}
with 
\begin{eqnarray}
    \mathcal{V}&=&S^3X \qquad \textrm{and}\qquad T_{\textrm{\tiny gr}}=\frac{1}{2\pi}\Bigg|\frac{\dot S}{S}+\frac{\dot X}{X}\Bigg|,
\end{eqnarray}
where we have imposed energy conservation of the effective energy-momentum tensor using the relation given in \ref{sec:divfree-tracefree entropy relation}. We emphasize that this choice applies only to the specific case of Szekeres II spacetimes. In the general case of Petrov type I spacetimes, it is not possible to prescribe a choice of factorization of the Bel-Robinson tensor, since this algebraic class contains spacetimes with a very broad range of geometric and physical properties, leading to a wide variety of factorizations of the Bel-Robinson tensor.

\subsection{Kinematic interpretation of the energy flux $q^a$}
In order to consider a Petrov type I spacetime as a small perturbation about a Petrov type D spacetime, $|\alpha|<<1$, a regime associated with $|q|<<|\rho-\bar\rho|$. The description of $q^a$ as an energy flux is not appropriate for a gravity dominated cosmological context.  Rather, a more meaningful physical interpretation can instead be interpreted kinematically as a peculiar velocity relating two 4-velocities $u^a$ and $\hat u^a$ by a boost \cite{najera2021pancakes,najera2021non, ellis2012relativistic, tsagas2010large} . 
\begin{eqnarray}
    \hat u^a &=& \gamma(u^a + v^a)\,, \quad \gamma = \sqrt{1-v^av_a}.
\end{eqnarray}
If one considers dust as the only source in the rest frame of $\hat u^a$, \textit{i.e.} $\hat T_{ab}= \hat\rho \hat u_a\hat u_b$, then it can be related to the Eulerian frame sources in Eq.~(\ref{EMEFE}), as
\begin{eqnarray}
    \hat\rho &\approx& \rho\, \quad \textrm{and}\quad q^a\approx \hat\rho v^a
\end{eqnarray}
if $v^a v_a<<1$. This approximation is precisely consistent in the regime $\alpha<<1$. Thus, Eq.~(\ref{entropy-B1-sources}) can be physically interpreted as having a peculiar velocity field on top of an inhomogeneous physical matter density which would enhance non-linearities and increase entropy.
\section{Conclusion}\label{concl}
We have examined the gravitational entropy of CET in Petrov type I spacetimes through the determination of the effective energy momentum tensors as factors that result from the non-unique algebraic decomposition of the  Bel-Robinson tensor.  This work extends the notion of CET entropy that was previously restricted to Petrov type D and N spacetimes for which this factorization is unique. We have examined the convergence of the non-unique factors of the  Bel-Robinson tensor in Petrov type I to the unique factors in Petrov types D and N. We found that, among the six factors in Petrov type I, two converge to the unique square root of Petrov type D when the Weyl scalars $\Psi_1$ and $\Psi_3$ vanish, while the remaining factors converge to the radiation factors of type D. The convergence to the correct limit suggests the viability of the factors in Petrov type I spacetime. 

We also obtained the gravitational states corresponding to each of these traceless factors. Surprisingly, each one of the factors obeys the same gravitational equation of state ($p_{\textrm{\tiny{gr}}}=\frac{1}{3}\rho_{\textrm{\tiny{gr}}}$). As an illustration of the framework, we considered the Szekeres class II solutions as a particular case of Petrov type I. For these solutions the four radiation factors have the same gravitational density and the remaining two Coulombian factors (associated to type D) differ at order $\mathcal{O}(\alpha^2)$. We calculated the entropy production for each factor and coupled the evolution to the Einstein field equations. We found that at leading order in $\alpha$ radiation states exhibit a higher entropy growth than in the Coulombic states, for which the first correction term due to $\alpha$ happens at $\mathcal{O}(\alpha^2)$. As such, for small $\alpha$, it allowed for a kinematic interpretation of the energy flux in terms of a peculiar velocity that enhances the entropy change. 

The CET entropy proposal is a promising candidate for a non-holographic coarse-grained description of gravitational entropy. An interesting future direction of this work would be to compare the results of entropy with other such gravitational entropy proposals. It would also be worthwhile to extend the analysis to general Petrov type I cosmological models, without the simplifying restrictions adopted in this work, in order to test the robustness of the associated gravitational arrow of time. A more theoretical aspect is to explore the factorization scheme in modified gravity scenarios that affect the Weyl tensor and to look into generic features of gravitational entropy in such scenarios. 
     
\section*{Acknowledgements}
We would like to thank Timothy Clifton for an insightful discussion. MS is supported by the {\it Agence Nationale de la Recherche} under the grant ANR-24-CE31-6963-01, and the French government under the France 2030 investment plan, as part of the Initiative d’Excellence d'Aix-Marseille Université -  A*MIDEX (AMX-19-IET-012). SN acknowledges support from a postdoctoral grant from Secretar\'ia de Ciencia, Humanidades, Tecnolog\'ia e Innovaci\'on SECIHTI. 

\appendix

\section{Relation between entropies of a traceless and an energy conserved effective energy momentum tensor}\label{sec:divfree-tracefree entropy relation}
Let $\{\rho_{_\textrm{\tiny{gr}}},p{_\textrm{\tiny{gr}}}\}$ denote the gravitational states of a traceless effective energy momentum $t_{ab}$ and $\{\tilde \rho{_\textrm{\tiny{gr}}},\tilde p{_\textrm{\tiny{gr}}}\}$ be the gravitational states of an effective energy momentum $\tilde t_{ab}$ with a non-zero trace. Owing to the freedom in determining the factors of the Bel-Robinson, we can relate these two by an arbitrary function $f$ as
\begin{eqnarray}
    \tilde t_{ab}&= t_{ab} + f g_{ab},
\end{eqnarray}
which gives us the relation 
\begin{eqnarray}
    \tilde {\rho}_{_\textrm{\tiny gr}} &= \rho_{_\textrm{\tiny gr}} -f, \qquad \tilde{p}{_\textrm{\tiny{gr}}} = p{_\textrm{\tiny{gr}}}+f.
\end{eqnarray}
Using Eq.~(\ref{Tds2}) we can relate the two entropies by keeping the same definition of the gravitational temperature as 
\begin{eqnarray}\label{ent-rel}
    T\dot{\tilde s} &= T\dot s-\dot f\mathcal{V}.
\end{eqnarray}
If we impose energy conservation on $\tilde t_{ab}$, \textit{i.e} $u_a\nabla_b \tilde t^{ab}\equiv u_a J^a=0$ then, $f$ can be fixed as
\begin{eqnarray}\label{f}
    u_a\nabla_b \left(t^{ab} +f g^{ab}\right) &=0\qquad \Rightarrow\qquad \dot f= -u_a J^a.
\end{eqnarray}
Using Eq.~(\ref{f}) in Eq.~(\ref{ent-rel}), we get the relation between the gravitational entropies of a traceless effective energy momentum tensor and an energy consverved effective energy-momentum tensor as
\begin{eqnarray}\label{ent-rel-2}
    T\dot{\tilde s} &= T\dot s+ \mathcal{V}u_a J^a\,.
\end{eqnarray}

Moreover, the integrability of the entropy associated with either a traceless effective energy momentum tensor or an energy-conserved effective energy momentum tensor, cf. \ref{sec:divfree-tracefree entropy relation} would ensure the integrability of the other if the following condition is satisfied
\begin{eqnarray}
    \partial_b\left(\frac{\V}{\Tgr}\right)\partial_a f - \partial_a\left(\frac{\V}{\Tgr}\right)\partial_b f &=0.
\end{eqnarray}
\section{Gravitational energy flux and gravitational anisotropic pressure of the traceless factors of the Bel-Robinson Tensor}\label{AppB}
For completeness, we calculate $q_{a}^{\textrm{\tiny{gr}}}$ and $\Pi_{a}^{\textrm{\tiny{gr}}}$ for each of the factors here. 

For the factor $A_1^{ab}$, the gravitational state variables are\begin{eqnarray}
    q^{{\textrm{\tiny gr}}}_a&=&0, \quad \quad \Pi^{{\textrm{\tiny gr}}}_{ab} = 2|\Psi_2|S^2\; \textrm{diag}\left(0,-2X^2,1,1\right)_{ab}.\end{eqnarray}

For $B_1^{ab}$, the non-vanishing components of the gravitational energy flux are
\begin{eqnarray}
q^{{\textrm{\tiny gr}}}_1&=&-\frac{8SX}{3|\Psi_2|}\left(|\Psi_1|^2-|\Psi_3|^2\right),\\
q^{{\textrm{\tiny gr}}}_2&=& -2S|\Psi_2|\left[\frac{1}{\bar\Psi_2}(\bar\Psi_1+\bar\Psi_3)+\frac{1}{\Psi_2}(\Psi_1+\Psi_3)\right],\\
q^{{\textrm{\tiny gr}}}_3&=& -2iS|\Psi_2|\left[\frac{1}{\bar\Psi_2}(\bar\Psi_1-\bar\Psi_3)-\frac{1}{\Psi_2}(\Psi_1-\Psi_3)\right],
\end{eqnarray}
while the non-vanishing components of the gravitational anisotropic pressure tensor are
\begin{eqnarray}
\Pi^{{\textrm{\tiny gr}}}_{11} &= S^2 X^2 |\Psi_2|\left[\left(\frac{4|\Psi_1|}{3|\Psi_2|}\right)^2+\left(\frac{4|\Psi_3|}{3|\Psi_2|}\right)^2-4\right],\\
\Pi^{{\textrm{\tiny gr}}}_{12} &= 2S^2X |\Psi_2|\left[(\Psi_1-\Psi_3)\frac{\bar\Psi_2}{|\Psi_2|^2}+(\bar\Psi_1-\bar\Psi_3)\frac{\Psi_2}{|\Psi_2|^2}\right],\\
\Pi^{{\textrm{\tiny gr}}}_{13} &= 2iS^2X |\Psi_2|\left[-(\Psi_1+\Psi_3)\frac{\bar\Psi_2}{|\Psi_2|^2}+(\bar\Psi_1+\bar\Psi_3)\frac{\Psi_2}{|\Psi_2|^2}\right],\\
\Pi^{{\textrm{\tiny gr}}}_{22} &= 2S^2 |\Psi_2|\left[1-\frac{4}{9|\Psi_2|^2}\left(|\Psi_1|^2+|\Psi_3|^2\right)+\frac{4}{3|\Psi_2|^2}\left(\Psi_1\bar\Psi_3+\bar\Psi_1\Psi_3\right)\right]\\
\Pi^{{\textrm{\tiny gr}}}_{23} &= \frac{8iS^2}{3}|\Psi_2|\left[-\frac{\Psi_1\bar\Psi_3}{|\Psi_2|^2}+\frac{\bar\Psi_1\Psi_3}{|\Psi_2|^2}\right],\\
\Pi^{{\textrm{\tiny gr}}}_{33} &=  2S^2 |\Psi_2|\left[1-\frac{4}{9|\Psi_2|^2}\left(|\Psi_1|^2+|\Psi_3|^2\right)-\frac{4}{3|\Psi_2|^2}\left(\Psi_1\bar\Psi_3+\bar\Psi_1\Psi_3\right)\right].\nonumber\\
&
\end{eqnarray}

For $A_{ab}^\pm$, the non-vanishing components of the gravitational energy flux are
\begin{eqnarray}
q^{{\textrm{\tiny gr}}}_1&=&SX(\Xi\pm3\Psi_2),\quad q^{{\textrm{\tiny gr}}}_2=\mp 4S\;\textrm{Re}(\Psi_1), \quad q^{{\textrm{\tiny gr}}}_3=\mp 4S\;\textrm{Im}(\Psi_1),
\end{eqnarray}
while the non-vanishing components of the gravitational anisotropic pressure are
\begin{eqnarray}
\Pi^{{\textrm{\tiny gr}}}_{11} &=&\frac{2S^2X^2}{3(\Xi\pm3\Psi_2)}\left[\Xi^2\pm 6\Xi\Psi_2+9\Psi_2^2-8|\Psi_1|^2\right],\\
\Pi^{{\textrm{\tiny gr}}}_{12} &=& \mp 4S^2X\;\textrm{Re}(\Psi_1),\\
\Pi^{{\textrm{\tiny gr}}}_{13} &=& \mp 4S^2X\;\textrm{Im}(\Psi_1),\\
\Pi^{{\textrm{\tiny gr}}}_{22} &=& \Pi^{{\textrm{\tiny gr}}}_{33} = -\frac{S^2}{3(\Xi\pm3\Psi_2)}\left[\Xi^2\pm 6\Xi\Psi_2+9\Psi_2^2-8|\Psi_1|^2\right].
\end{eqnarray}

For $B_{ab}^\pm$, the non-vanishing components of the gravitational heat flux are
\begin{eqnarray}
q^{{\textrm{\tiny gr}}}_1&=&-SX(\Xi\pm3\Psi_2),\quad q^{{\textrm{\tiny gr}}}_2=\mp 4S\;\textrm{Re}(\Psi_3), \quad q^{{\textrm{\tiny gr}}}_3=\pm 4S\;\textrm{Im}(\Psi_3),
\end{eqnarray}
while the non-vanishing components of the gravitational anisotropic pressure are
\begin{eqnarray}
\Pi^{{\textrm{\tiny gr}}}_{11} &=&\frac{2S^2X^2}{3(\Xi\pm3\Psi_2)}\left[\Xi^2\pm 6\Xi\Psi_2+9\Psi_2^2-8|\Psi_3|^2\right],\\
\Pi^{{\textrm{\tiny gr}}}_{12} &=& \pm 4S^2X\;\textrm{Re}(\Psi_3),\\
\Pi^{{\textrm{\tiny gr}}}_{13} &=& \mp 4S^2X\;\textrm{Im}(\Psi_3),\\
\Pi^{{\textrm{\tiny gr}}}_{22} &=& \Pi^{{\textrm{\tiny gr}}}_{33} = -\frac{S^2}{3(\Xi\pm3\Psi_2)}\left[\Xi^2\pm 6\Xi\Psi_2+9\Psi_2^2-8|\Psi_3|^2\right].
\end{eqnarray}

\section*{References}
\bibliographystyle{iopart-num}
\bibliography{main}

\end{document}